\documentclass[12pt, draftcls, onecolumn]{IEEEtran}
\usepackage{pdfsync,srcltx}
\usepackage{amsmath}
\usepackage{amsfonts}
\usepackage{amssymb}
\usepackage{graphicx,subfigure,psfrag}
\usepackage{url,ifthen}
\usepackage{pmat}
\usepackage{cite}
\usepackage[ieee]{jphmacros2e}

\usepackage[norefs,nocites]{refcheck}
\makeatletter\let\NAT@parse\undefined\makeatother 

\newcommand{\R}{\mathbb{R}}

\newcommand{\N}{\mathbb{N}}
\newcommand{\Z}{\mathbf{Z}}

\newcommand{\G}{\mathbf{G}}
\newcommand{\V}{\mathbf{V}}
\renewcommand{\E}{\mathbf{E}}
\newcommand{\scr}{\mathcal}
\newcommand{\comment}[1]{}

\newcommand{\PaperORReport}{Report}

\allowdisplaybreaks

\ifthenelse{\equal{\PaperORReport}{Paper}}{
\title{On achieving size-independent stability margin of vehicular lattice formations with distributed control}
 }
 {
 \title{\LARGE On achieving size-independent stability margin of vehicular lattice formations with distributed control}
 }
\author{He Hao,
        Prabir Barooah
	\thanks{He Hao and Prabir Barooah are with Department of
          Mechanical and Aerospace Engineering, University of Florida,
          Gainesville, FL 32611, USA. This work was supported by the National Science Foundation through Grant CNS-0931885 and ECCS-0925534.}}

\begin{document}
\maketitle

\begin{abstract}
  We study the stability margin of a vehicular formation with
  distributed control, in which the control at each vehicle only
  depends on the information from its neighbors in an information
  graph. We consider a D-dimensional lattice as information graph, of
  which the 1-D platoon is a special case. The stability margin is
  measured by the real part of the least stable eigenvalue of the
  closed-loop state matrix, which quantifies the rate of decay of
  initial errors.  In~\cite{PB_PM_JH_TAC:09}, it was shown that with
  symmetric control, in which two neighbors put equal weight on
  information received from each other, the stability margin of a 1-D
  vehicular platoon decays to $0$ as $O(1/N^2)$, where $N$ is the
  number of vehicles. Moreover, a perturbation analysis was used to
  show that with vanishingly small amount of asymmetry in the control
  gains, the stability margin scaling can be improved to $O(1/N)$. In
  this paper, we show that, with judicious choice of non-vanishing
  asymmetry in control, the stability margin of the closed loop can be
  bounded away from zero uniformly in $N$. Asymmetry in control gains
  thus makes the control architecture highly scalable. The results are
  also generalized to $D$-dimensional lattice information graphs that
  were studied in~\cite{HH_PB_PM_TAC:11}, and the correspondingly
  stronger conclusions than those derived in~\cite{HH_PB_PM_TAC:11}
  are obtained. In addition, we show that the size-independent
  stability margin can be achieved with relative position and
  \emph{relative velocity} (RPRV) feedback as well as relative
  position and \emph{absolute velocity} (RPAV) feedback, while the
  analysis in~\cite{PB_PM_JH_TAC:09,HH_PB_PM_TAC:11} was only for the
  RPAV case.
\end{abstract}


\section{Introduction}
We study cooperative control of a large vehicular
formation with distributed control. The vehicles are modeled as double
integrators, and the control action at each vehicle is computed based
on information from its neighbors, where the neighbor relationship is
characterized by a lattice information graph. The control objective is to
make the vehicular formation track a constant-velocity type desired trajectory while
maintaining pre-specified constant separation among neighbors. The desired trajectory of the
entire vehicular formation is given in terms of trajectories of a set
of fictitious reference vehicles.

The problem of distributed control for multi-agent coordination is
relevant to many applications such as automated highway system,
collective behavior of bird flocks and animal swarms, and formation
flying of unmanned aerial and ground vehicles for surveillance,
reconnaissance and rescue,
etc.~\cite{JH_MT_PV_CSM:94,okubo_flock,DJ_WB_MP_EW_AIAA:02,
  SD_PP_IJES:10,das2002framework,tanner2004leader}. A typical issue
faced in distributed control is that as the number of agents
increases, the performance (stability margin and sensitivity to external disturbances) of the closed loop degrades.  Several recent papers have studied the scaling of performance  of vehicle formations as a function of the number of vehicles.  The references~\cite{PB_PM_JH_TAC:09,HH_PB_PM_TAC:11} have studied the
scaling of the stability margin of $D$-dimensional lattice
formations. The stability margin is defined as the absolute value of
the real part of the least stable eigenvalue of the closed loop. The
stability margin characterizes the rate at which initial errors
decay. The
references~\cite{stringstability:96,Seiler_disturb_TAC:04,bamjovmitCDC08,illposed,veerman2010asymmetric}
have examined the sensitivity of 1-dimensional platoons to external
disturbances. However, among papers that examined sensitivity to
disturbance, to the best of our knowledge only
\cite{veerman2010asymmetric} has considered asymmetric control, the rest
are limited to symmetric control. The control is called symmetric if between two
neighboring vehicles $i$ and $j$, the weight $i$ puts on the
information from $j$ is the same as the weight $j$ puts on the
information from $i$.

In previous works on 1-D vehicular platoons, two types of feedback
are respectively considered: relative position absolute velocity
(RPAV) feedback~\cite{illposed,PB_PM_JH_TAC:09} and relative position
relative velocity (RPRV)
feedback~\cite{bamjovmitCDC08,veerman2010asymmetric,HH_PB_CDC:10}. With
symmetric control, the stability margin of the vehicular platoon decays to $0$ as $O(1/N^2)$ in both types of
feedback. This result for RPAV feedback was shown
in~\cite{PB_PM_JH_TAC:09}, and for RPRV feedback was shown
in~\cite{HH_PB_CDC:10}.  The loss of stability margin with symmetric
control has also been recognized by other
researchers~\cite{illposed,veerman_automated}. Asymmetric control in
the RPAV case was examined in~\cite{PB_PM_JH_TAC:09,HH_PB_PM_TAC:11},
where it was also shown that with vanishingly small asymmetry in
the control gains, the stability margin can be improved to
$O(1/N)$. Similar conclusions are also obtained for a vehicle
formation with a $D$-dimensional lattice as its information
graph~\cite{HH_PB_PM_TAC:11} - that decay of stability margin can be
improved with asymmetry. In case of RPRV feedback, a similar
improvement to $O(1/N)$ with asymmetry was shown in~\cite{HH_PB_CDC:10}, where only
the relative velocity feedback gains were made asymmetric. The analyses
in~\cite{PB_PM_JH_TAC:09,HH_PB_CDC:10,HH_PB_PM_TAC:11} were based on a
partial differential equation (PDE) approximation of the closed loop
dynamics and a perturbation method; the latter limited the results to
only vanishingly small asymmetry.

In this paper we provide a stronger result on the stability margin
with asymmetric control by avoiding the perturbation
analysis of the aforementioned papers. We also  avoid the PDE
approximation and analyze the state space model directly. In particular, we show that with judicious choice of asymmetry in the control, the stability
margin of the vehicular formation can be uniformly bounded away from
$0$ (independent of $N$) and derive a closed-form formula for the
lower bound. This result makes it possible to
design the control gains so that the stability margin of the system
satisfies a pre-specified value irrespective of how many vehicles
are in the formation. We also generalize the result to formations
with $D$-dimensional information graphs, and show that a similar,
size-independent stability margin can be obtained by using asymmetry
in the control gains. These results are established for both RPAV and
RPRV feedbacks.  

The focus of this paper is on the stability margin, which is related
to exponential stability of the closed loop system. A related concept
is that of ``string stability''~\cite{SwaroopHedrick_stringstability_TAC:96}.
String stability is usually interpreted as the system's sensitivity to
external disturbances;
see~\cite{zhang1999using,Seiler_disturb_TAC:04,Middleton_TAC:09,SD_PP_IJES:10}
and references therein. We do not study sensitivity to
external disturbances in this paper. 

For ease of description, we first present the problem statement and
main result for a vehicular formation with $1$-dimensional information
graph (i.e. the vehicular platoon) in Section~\ref{sec:problem}.
Analysis of the stability margin and numerical verification appear in
Section~\ref{sec:conv_rate}.  The extension of the result to a vehicular
formation with $D$-dimensional lattice information graph is presented in Section~\ref{sec:sm_nD}. The paper ends with a summary in Section~\ref{sec:conc}.

\section{Problem statement and result for 1-D platoon}\label{sec:problem}
\subsection{Problem statement}\label{sec:statement}
\begin{figure}
	  \psfrag{O}{$O$}
	  \psfrag{o}{$0$}
	  \psfrag{l}{$1$}
	  \psfrag{X}{$X$}
	  \psfrag{x}{$x$}
	  \psfrag{d1}{\scriptsize$\Delta_{(0,1)}$}
	  \psfrag{d2}{\scriptsize$\Delta_{(N-1,N)}$}
	   \psfrag{v}{$v^{*}\;t$}
	   \psfrag{d}{\scriptsize $1/N$}
	   \psfrag{0}{\scriptsize $0$}
	   \psfrag{1}{\scriptsize $1$}
	   \psfrag{N-1}{\scriptsize $N-1$}
	   \psfrag{N}{\scriptsize $N$}
	   \psfrag{Di}{\scriptsize Dirichlet}
	   \psfrag{Ne}{\scriptsize Neumann}
\centering
\ifthenelse{\equal{\PaperORReport}{Paper}}{
 \includegraphics[scale = 0.3]{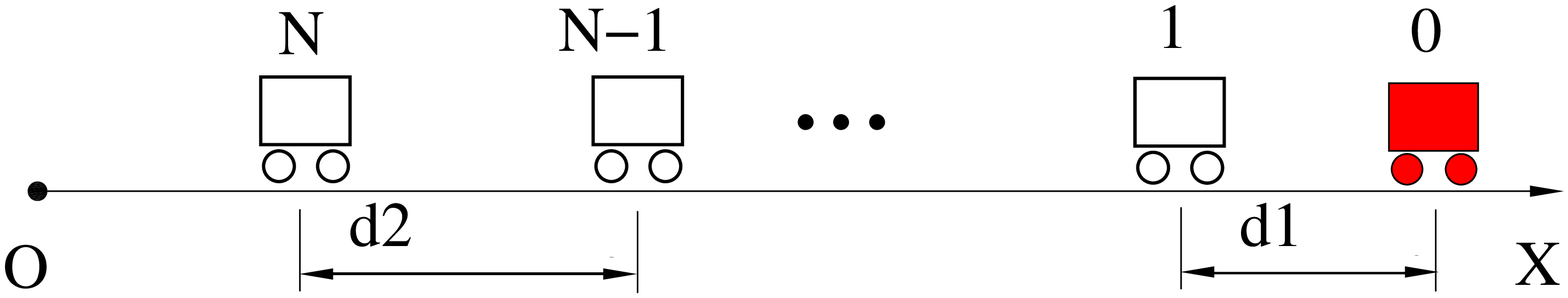} 
 }
 {
  \includegraphics[scale = 0.3]{fig1.eps} 
 }
\caption{Desired geometry of a vehicular platoon with $N$ vehicles and $1$ ``fictitious'' reference vehicle. The filled vehicle in the front of the  platoon represents the reference vehicle, it is denoted by index $0$. }
\label{fig:fig1}
\end{figure}

In this section we consider the formation control of $N$ homogeneous vehicles which are moving in 1-D Euclidean space, as shown in Figure~\ref{fig:fig1}. The position of the $i$-th vehicle is denoted by  $p_i \in \R$ and the dynamics of each vehicle are modeled as a double integrator:
\begin{align}\label{eq:vehicle-dynamics}
	\ddot{p}_i= u_i, \quad i\in \{1,2,\cdots,N\}, 
 \end{align}
where $u_i \in \R$ is the control input.  This is a commonly used model for vehicle dynamics in studying vehicular formations, which results from feedback linearization of  non-linear vehicle dynamics~\cite{darbha1994comparison,feedback_linearization}.

The control objective is that vehicles maintain a desired formation geometry while following a constant-velocity type desired trajectory. The desired geometry of the formation is specified by the \emph{desired gaps} $\Delta_{(i-1,i)}$ for $i\in\{1,\cdots,N\}$, where $\Delta_{(i-1,i)}$ is the desired value of $p_{i-1}(t) - p_{i}(t)$. The desired inter-vehicular gaps $\Delta_{(i-1,i)}$'s are positive constants and they have to be specified in a mutually consistent fashion, i.e. $\Delta_{(i,k)}=\Delta_{(i,j)} +\Delta_{(j,k)}$ for every triple $(i, j, k)$ where $i \leq j \leq k$. The desired trajectory of the platoon is provided in terms of a \emph{fictitious} reference vehicle with index $0$, whose trajectory is given by $p^*_0(t) = v^*t + c_0$ for some constants $v^*,c_0$, where $v^*$ is the cruise velocity of the formation.  The desired trajectory of the $i$-th vehicle, $p^*_i(t)$, is given by $p^*_i(t) =p^*_0(t)-\Delta_{(0,i)}=p^*_0(t)-\sum_{j=1}^i\Delta_{(j-1,j)}$.

We consider the following \emph{distributed} control laws.

1) \emph{Relative position and absolute velocity  (RPAV) feedback}: the control action at the $i$-th
vehicle depends on the relative position measurements with its two 
neighbors (one on either side), its own velocity, and the desired velocity
$v^*$:
\begin{align}\label{eq:control-law}
	u_{i} = &-k_{i}^{f}(p_{i}-p_{i-1}+\Delta_{(i-1,i)}) -k_{i}^{b}(p_{i}-p_{i+1}-\Delta_{(i,i+1)})\notag \\ &-b_{i}(\dot{p}_{i}-v^*),  \quad i \in \{1,\cdots,N-1 \}, \notag \\
	u_{N} = &-k_{N}^{f}(p_{N}-p_{N-1}+\Delta_{(N-1,N)})-b_{N}(\dot{p}_{N}-v^*),
\end{align}
where  $k_{i}^{f}, k_{i}^{b}$ are the front and back position gains and $b_{i}$ is the velocity gain.

2) \emph{Relative position and relative velocity (RPRV) feedback}: the control action at the $i$-th
vehicle depends on the relative position and relative velocity measurements with its nearest 
neighbors in the platoon:
\begin{align}\label{eq:control-law1}
	u_{i} = &-k_{i}^{f}(p_{i}-p_{i-1}+\Delta_{(i-1,i)}) -k_{i}^{b}(p_{i}-p_{i+1}-\Delta_{(i,i+1)})\notag \\
	&-b_{i}^{f}(\dot{p}_{i}-\dot{p}_{i-1}) -b_{i}^{b}(\dot{p}_{i}-\dot{p}_{i+1}), \quad i \in \{1,\cdots,N-1 \}, \notag \\
	u_{N} = &-k_{N}^{f}(p_{N}-p_{N-1}+\Delta_{(N-1,N)})-b^f_{N}(\dot{p}_{N}-\dot{p}_{N-1}), 
\end{align}
where  $k_{i}^{f}, k_{i}^{b}$ (respectively, $b_{i}^{f}, b_{i}^{b}$) are the front and back position (respectively, velocity) gains of the $i$-th vehicle. 

In the RPRV feedback case, vehicle $i$ must be provided (a-priori)
the desired gaps with its two neighbors. In the RPAV feedback, it must
be provided with additional information: the formation's desired velocity $v^*$. 
The closed-loop dynamics with RPAV (resp., RPRV) feedback, in terms
of the tracking errors $\tilde{p}_i  \eqdef p_i-p_i^* $, can now be expressed as: 
\begin{align}\label{eq:ss}
	\dot{x}  &= A^{(\text{\tiny{RPAV}})}x, &\quad (\text{resp.}) \quad \dot{x}  &= A^{(\text{\tiny{RPRV}})}x,
\end{align}
where the state vector is defined as $x \eqdef
[\tilde{p}_1,\dot{\tilde{p}}_1,\cdots,\tilde{p}_N,\dot{\tilde{p}}_N] \in
\R^{2N}$, and the state matrix $A^{(.)}$ depends on the control gains
but not on the desired gaps or desired velocity.  

\begin{definition}\label{def:main-def}
	The \emph{stability margin} $S^{(\text{\tiny{RPAV}})}$ (respectively, $S^{(\text{\tiny{RPRV}})}$) of the closed-loop system~\eqref{eq:ss} is defined as the absolute value of the real part of the least stable eigenvalue of $A^{(\text{\tiny{RPAV}})}$ (respectively, $A^{(\text{\tiny{RPRV}})}$). The control law~\eqref{eq:control-law} (respectively, \eqref{eq:control-law1}) is \emph{symmetric} if each
vehicle uses the same front and back control gains: $k_{i}^{f}=k_{i}^{b}=k_0, b_{i}=b_0$
(respectively, $k_{i}^{f}=k_{i}^{b}=k_0, b_{i}^{f}=b_{i}^{b}=b_0$), for all $i \in \{1, 2,\cdots, N-1 \}$, where $k_0, b_0$ are positive constants. \frqed
\end{definition}

In this paper, we consider the following \emph{asymmetric} control gains
\begin{align}\label{eq:control_gains}
\text{RPAV feedback:  }  k^f_i =(1+\epsilon)k_0,  \ \   k^b_i=(1-\epsilon)k_0, \ \  b_i =b_0.
\end{align}
\begin{align}\label{eq:control_gains1}
\text{RPRV feedback: }  \begin{split} k^f_i =(1+\epsilon)k_0,\quad
  k^b_i=(1-\epsilon)k_0, \\ b^f_i=(1+\epsilon)b_0,\quad
  b^b_i=(1-\epsilon)b_0, 
\end{split} \quad \quad \quad \quad 
\end{align}
where $\epsilon \in [0,1)$ denotes the amount of
asymmetry; $\epsilon=0$ corresponds to symmetric
control. The design for the RPAV case is inspired
by~\cite{PB_PM_JH_TAC:09,HH_PB_PM_TAC:11}. The control gains given
in~\eqref{eq:control_gains} and~\eqref{eq:control_gains1} are
homogeneous in the sense that they do not vary with $i$. The reason we
only consider homogeneous control gains is that heterogeneity has little effect
on the scaling of stability margin, see~\cite{HH_PB_CDC:10} for a proof for 1-D platoon. The proof  for vehicular formation with general graphs is given in Lemma~\ref{lem:hete}, which is provided in the appendix.

The following proposition summaries the results in~\cite{PB_PM_JH_TAC:09, HH_PB_CDC:10}.
\begin{proposition}\label{prop:summary} Consider an $N$-vehicle platoon with closed loop dynamics~\eqref{eq:ss}.
        \begin{enumerate}
        \item [1)] [Corollary 1 of~\cite{PB_PM_JH_TAC:09}, Theorem 1 of~\cite{HH_PB_CDC:10}] With symmetric control ($\epsilon=0$), both $S^{(\text{\tiny{RPAV}})}$ and $S^{(\text{\tiny{RPRV}})}$ are $O(\frac{1}{N^2}).$
      \item [2)] [Corollary 3 of~\cite{PB_PM_JH_TAC:09}] With the asymmetric control gains
         $k^f_i=k_0 (1+\epsilon)$, $k^b_i=k_0(1-\epsilon)$ and $b_i=b_0$, 
        the  stability margin of the platoon with RPAV feedback is $S^{(\text{\tiny{RPAV}})}  = O(\frac{\epsilon}{N})$. \footnote{The case considered in~\cite{PB_PM_JH_TAC:09} was that $|k^f_i - k_0| < \epsilon$ , $|k^b_i - k_0| <
  \epsilon$. It is straightforward, however, to re-derive the results if the
  constraints on the gains are changed to the form used here:
  $|k^f_i - k_0|/k_0 < \epsilon$ , $|k^b_i - k_0|/k_0 <
  \epsilon$. In this paper we consider the latter case
  since it makes the analysis cleaner without changing the results
  of~\cite{PB_PM_JH_TAC:09} significantly.
} 
\item [3)] [Theorem 2 of~\cite{HH_PB_CDC:10}] With asymmetric control gains $k^f_i=k^b_i=k_0, b^f_i=b_0 (1+\epsilon),b^b_i=b_0(1-\epsilon)$, 
        the  stability margin of the platoon with RPRV feedback is $S^{(\text{\tiny{RPRV}})}  = O(\frac{\epsilon}{N})$. 
        \end{enumerate}
        Statements (2) and (3) hold in the limit $\epsilon \to 0$
        and $N \to \infty$. \frqed
\end{proposition}

Proposition~\ref{prop:summary} shows that with symmetric control, the
stability margin decays to $0$ as $O(1/N^2)$, irrespective of the type
of feedback we used. However, in the case of RPAV feedback, with
vanishingly small amount of asymmetry in the position gains, the
stability margin of the system can be improved to $O(1/N)$. The same
$O(1/N)$ trend can be achieved for the case of RPRV feedback with
vanishingly small asymmetry in the velocity gains alone while the
position gains are held symmetric. The
design~\eqref{eq:control_gains1} was not considered
in~\cite{HH_PB_CDC:10}. Since the results
in~\cite{PB_PM_JH_TAC:09,HH_PB_CDC:10} were obtained with
a perturbation analysis, these results are applicable only when the
amount of asymmetry is  vanishingly small.

  The following theorem is the main result of this paper, whose proof
 and numerical corroboration are given in Section~\ref{sec:conv_rate}.

\begin{theorem}\label{thm:sm_bound} With the control gains given
in~\eqref{eq:control_gains} and~\eqref{eq:control_gains1} respectively, for any fixed $\epsilon \in (0, \;1)$, the closed loop is exponentially stable and the stability margin of the vehicular platoon  is  bounded away from $0$ uniformly in $N$. Specifically,
\begin{align}
S^{(\text{\tiny{RPAV}})} \geq & \frac{\Re \Big (b_0-\sqrt{b_0^2-8k_0(1-\sqrt{1-\epsilon^2})}\Big )}{2}, \label{eq:sm_bound}\\
S^{(\text{\tiny{RPRV}})} \geq & \min \Big \{ b_0(1-\sqrt{1-\epsilon^2}), \  \frac{k_0}{b_0} \Big\}, \label{eq:sm_bound1}
\end{align} 
where $\Re (.)$ denotes the real part.   \frqed
\end{theorem}

\begin{remark}
  Comparing Theorem~\ref{thm:sm_bound}  with
  Proposition~\ref{prop:summary}, we observe the following: (1)  Even with an arbitrarily
  small (but fixed and non-vanishing) amount of asymmetry in the control gains, the
  stability margin of the system can be bounded away from zero
  \emph{uniformly in $N$}. This asymmetric design therefore makes the
  resulting control law highly scalable; it eliminates the degradation
  of stability margin with increasing $N$. (2) In case of the RPAV feedback,
  although the control law is the same as that analyzed
  in~\cite{PB_PM_JH_TAC:09}, the stronger conclusion we obtained -
  compared to that in~\cite{PB_PM_JH_TAC:09} - is due to the fact that
  our analysis does not rely on a perturbation-based technique that
  was used~\cite{PB_PM_JH_TAC:09}, which limited the analysis
  in~\cite{PB_PM_JH_TAC:09} to vanishingly small $\epsilon$. (3) For
  the RPRV feedback case, the stronger result compared to that
  in~\cite{HH_PB_CDC:10}, is obtained by putting equal asymmetry in
  both position and velocity gains, while~\cite{HH_PB_CDC:10} allowed
  asymmetry only in the velocity gain. In  addition, unlike~\cite{PB_PM_JH_TAC:09,HH_PB_CDC:10}, we do not use a PDE
  (partial differential equation) approximation to analyze the
  stability margin, but analyze the state-space model directly. 
  \frqed
\end{remark}

\section{Stability margin of the 1-D vehicular platoon}\label{sec:conv_rate}
With the control gains specified in~\eqref{eq:control_gains} and~\eqref{eq:control_gains1} respectively, it can be shown that the state matrices  can be expressed in the following forms,
\begin{align}\label{eq:matrix_relation}
	A^{(\text{\tiny{RPAV}})} = I_N \otimes A_1 + L^{(1)} \otimes A_2,\notag\\
	A^{(\text{\tiny{RPRV}})} = I_N \otimes A_3 + L^{(1)} \otimes A_4,
\end{align}
where $I_N$ is the $N \times N$ identity matrix,  $\otimes$ denotes the
Kronecker product, and
\begin{align}\label{eq:auxi_matrices}
	A_1 \eqdef \begin{bmatrix}
		0 & 1 \\
		0 & -b_0 \\
	       \end{bmatrix}, \quad
	A_2 \eqdef \begin{bmatrix}
		0 & 0 \\
		-k_0 & 0 \\
	       \end{bmatrix},\notag\\
	A_3 \eqdef \begin{bmatrix}
		0 & 1 \\
		0 & 0 \\
	       \end{bmatrix}, \quad
	A_4 \eqdef \begin{bmatrix}
		0 & 0 \\
		-k_0 & -b_0 \\
	       \end{bmatrix},
\end{align}
where $k_0>0,b_0>0$ are the nominal position and velocity gains
respectively, and
\begin{align}\label{eq:redu_laplacian}
	L^{(1)} \eqdef
		\begin{bmatrix}
		2 & -1+\epsilon& &  \\
		-1-\epsilon & 2 &  -1+\epsilon & \\
		 &\ddots \quad \ & \ddots \quad \ & \ddots  \quad \ \\
		&  -1-\epsilon & 2 &  -1+\epsilon\\
		&  &-1-\epsilon  & 1+\epsilon
	       \end{bmatrix}.
\end{align}
It follows from Theorem 3.1 of~\cite{yueh2008explicit} that the
eigenvalues of $L^{(1)}$ are given by
 \begin{align}\label{eq:lambda_formula1}
	 \lambda&=b+2c\rho \cos \theta,
\end{align}
if $\theta$ ($\theta \neq m\pi, m \in \mathbb{Z}$, $\mathbb{Z}$ being
the set of integers) is a solution to
\begin{align}\label{eq:formula1}
	\rho^N(ac \sin(N+1)\theta+(\gamma \delta-\alpha \beta)\sin(N-1)\theta \notag \\-c \rho(\gamma+\delta)\sin N\theta)-(c\alpha \rho^{2N}+a\beta) \sin \theta=0,
\end{align}
where $a=-1-\epsilon, b=2, c=-1+\epsilon, \alpha=\beta=\gamma=0,
\delta=-1+\epsilon,\rho=\sqrt{(-1-\epsilon)/(-1+\epsilon)}$. Eq.~\eqref{eq:lambda_formula1} and~\eqref{eq:formula1} can now be simplified to
 \begin{align}\label{eq:lambda_formula}
	 \lambda_\ell&=2-2\sqrt{1-\epsilon^2} \cos \theta_\ell, \quad \ell \in\{1,2,\cdots, N\},
\end{align}
where $\epsilon \in (0,1)$ and $\theta_\ell$ is the $\ell$-th root of the following equation
\begin{align}\label{eq:formula}
	 \sqrt{\frac{1+\epsilon}{1-\epsilon}} \sin(N+1)\theta=\sin
         N\theta.
\end{align}
From~\eqref{eq:lambda_formula}, we see that the eigenvalues of
$L^{(1)}$ are real and positive, and moreover, $0 < \lambda_1= 2 - 2\sqrt{1 - \epsilon^2} \cos \theta_1 < \lambda_2 <\dots < \lambda_N=2 - 2\sqrt{1 - \epsilon^2} \cos \theta_N$, where $\theta_1 \in (\frac{\pi}{2(N+1)},\frac{3\pi}{2(N+1)}), \theta_N \in (\frac{(2N-1)\pi}{2(N+1)},\frac{(2N+1)\pi}{2(N+1)})$ are the solutions to~\eqref{eq:formula}. To see why, first notice that we only need
consider the roots of~\eqref{eq:formula} in the open interval
$(0,2\pi)$, in which there are $2N$ nontrivial isolated roots.
\ifthenelse{\equal{\PaperORReport}{Paper}}{}{See Figure~\ref{fig:loc_root} for an example.} The
roots located in $\R \setminus (0,2\pi)$ are $2m\pi$ ($m \in
\mathbb{Z}$) distance away from those in $(0,2\pi)$. Moreover, if
$\theta_0 \in (0,2\pi)$ is a solution of~\eqref{eq:formula}, then
$2\pi-\theta_0$ is also a solution. Therefore, we can restrict the
domain of analysis to $(0,\pi)$, in which there are $N$ isolated
roots. The ordering of the eigenvalues follows from $\cos \theta$ being
a decreasing function in $(0, \pi)$.  It is straightforward to show from graphical solution of~\eqref{eq:formula}
that the $\ell$-th root $\theta_\ell$ is in the open interval
$(\frac{(2\ell-1)\pi}{2(N+1)},\frac{(2\ell+1)\pi}{2(N+1)})$. We now
present a formula for the stability margin of the vehicular platoon in terms of the  eigenvalues of
$L^{(1)}$.
\ifthenelse{\equal{\PaperORReport}{Paper}}{ }{
\begin{figure}
	\psfrag{2pi}{$2\pi$}    	
	\psfrag{pi}{$\pi$}
	\psfrag{theta}{\quad $\theta$}
	\psfrag{sin1}{\scriptsize$\sqrt{(1+\epsilon)/(1-\epsilon)} \sin( (N+1)\theta)$}
	\psfrag{sin2}{\scriptsize $ \sin (N\theta)$}
\centering
\includegraphics[scale = 0.5]{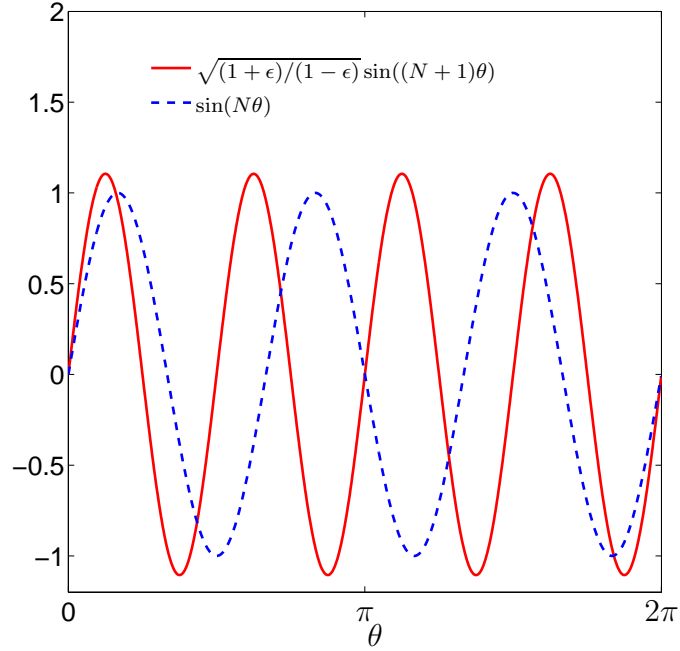}
\caption{Graphical solution $\theta$ of $\sqrt{(1+\epsilon)/(1-\epsilon)} \sin ( (N+1)\theta) = \sin (N\theta)$ with $\epsilon=0.1$ and $N=3$.}
\label{fig:loc_root}
\end{figure}
}

\begin{lemma}\label{lem:eig_relation}
	With the control gains given in~\eqref{eq:control_gains}
        and~\eqref{eq:control_gains1} respectively, and $0<\epsilon
        <1$, the stability margin of the vehicular platoon is 
\begin{align*}
	S^{(\text{\tiny{RPAV}})} &=
	\begin{cases}
		\frac{b_0}{2},& \text{if } \lambda_1 \geq b_0^2/4k_0, \\ 
		\frac{ b_0 - \sqrt{b_0^2-4k_0\lambda_1}}{2},&
                \text{otherwise,}
	\end{cases} \notag\\
	S^{(\text{\tiny{RPRV}})}&=
	\begin{cases}
		\frac{b_0 \lambda_1}{2},& \text{if } \lambda_N \leq 4k_0/b_0^2, \\ 
		\frac{2k_0}{b_0+\sqrt{b_0^2-4k_0/\lambda_N}}, & \text{if } \lambda_1 \geq 4k_0/b_0^2, \\
		\min \Big \{ \frac{b_0 \lambda_1}{2}, \frac{2k_0}{b_0+\sqrt{b_0^2-4k_0/\lambda_N}} \Big\},& \text{otherwise,}
	\end{cases}
\end{align*}
where $\lambda_1$ and  $\lambda_N$ are the smallest and largest eigenvalues of $L^{(1)}$ respectively. \frqed
\end{lemma}
\begin{proof-lemma}{\ref{lem:eig_relation}}
Our proof follows a similar line of attack as of~\cite{veerman2005flocks}. From Schur's triangularization theorem, there exists an unitary matrix $U$ such that
\begin{align*}
 	U^{-1} L^{(1)} U= L_u,
\end{align*}
where $L_u$ is an upper-triangular matrix whose diagonal entries are the eigenvalues $\lambda_\ell$ of $L^{(1)}$. We first consider the RPAV feedback case. We do a similarity transformation on matrix $A^{(\text{\tiny{RPAV}})}$.
 \begin{align*}
 	\bar{A}^{(\text{\tiny{RPAV}})} &\eqdef (U^{-1} \otimes I_2) A^{(\text{\tiny{RPAV}})} (U \otimes I_2)\\
	&=(U^{-1} \otimes I_2) (I_N \otimes A_1 + L^{(1)} \otimes A_2) (U \otimes I_2)\\
	&= I_N \otimes A_1+ L_u \otimes A_2.
\end{align*}
It is a block upper-triangular matrix, and the block on each diagonal is $A_1+\lambda_{\ell} A_2$, where $\lambda_{\ell} \in \sigma(L^{(1)})$, and $\sigma(\cdot)$ denotes the spectrum (the set of eigenvalues). Since similarity transformation preserves eigenvalues, and the eigenvalues of a block upper-triangular matrix are the union of eigenvalues of each block on the diagonal, we have
\begin{align}\label{eq:eigen_relation}
	\sigma(A^{(\text{\tiny{RPAV}})}) = \sigma(\bar{A}^{(\text{\tiny{RPAV}})})  &= \bigcup_{\lambda_{\ell} \in \sigma(L^{(1)})} \{ \sigma (A_1+\lambda_{\ell} A_2)\}\notag\\	&= \bigcup_{\lambda_{\ell} \in \sigma(L^{(1)})} \Big \{ \sigma  \begin{bmatrix}
		0 & 1 \\ -k_0\lambda_\ell & -b_0 	       \end{bmatrix}
	       \Big \}.
\end{align}
It follows now that the eigenvalues of $A^{(\text{\tiny{RPAV}})}$ are
the roots of the characteristic equation  $s^2+b_0 s + k_0 \lambda_{\ell}  =0$. For each $\ell \in \{1,2,\cdots, N\}$, the two roots are
 \begin{align}\label{eq:eigenvalue_pos_vel}
	 s_\ell^{\pm} = \frac{- b_0 \pm \sqrt{b_0^2-4k_0\lambda_{\ell}}}{2}.
 \end{align}
 The root closer to the imaginary axis is denoted by
 $s_\ell^+$, and is called the \emph{less stable} eigenvalue between
 the two. The \emph{least stable} eigenvalue is the one closet to the
 imaginary axis among them, it is denoted by $s_{\min}$. It follows
 from Definition~\ref{def:main-def} that $S = |\Re(s_\mrm{min})|$.

\ifthenelse{\equal{\PaperORReport}{Paper}}{
Depending on the  discriminant in~\eqref{eq:eigenvalue_pos_vel}, there are two cases to analyze:
(1) If $\lambda_1 \geq b_0^2/4k_0$, due to $\lambda_1<\cdots<\lambda_N$, we have the discriminant in~\eqref{eq:eigenvalue_pos_vel} for each $\ell$ is non-positive, which yields $S^{(\text{\tiny{RPAV}})}=|\Re(s_{\min})|=\frac{b_0}{2}$.
(2) Otherwise, the less stable eigenvalues are $s_\ell^+
= \frac{1}{2} (-b_0 + \sqrt{b_0^2-4k_0\lambda_\ell})$, which may be complex for some $\ell>1$. 
The least stable eigenvalue is obtained by setting
$\lambda_\ell=\lambda_1$, so that $S^{(\text{\tiny{RPAV}})}=|\Re(s_{\min})|=\frac{1}{2} (b_0 - \sqrt{b_0^2-4k_0\lambda_1})$.

The result on the stability margin of the platoon with RPRV feedback
follows by the same procedures as above, and is provided
in~\cite{HH_PB_TAC_arxiv:11}. \frQED
}{
Depending on the  discriminant in~\eqref{eq:eigenvalue_pos_vel}, there are two cases to analyze:
\begin{enumerate}
	\item If $\lambda_1 \geq b_0^2/4k_0$, due to $\lambda_1<\cdots<\lambda_N$, we have the discriminant in~\eqref{eq:eigenvalue_pos_vel} for each $\ell$ is non-positive, which yields $S^{(\text{\tiny{RPAV}})}=|\Re(s_{\min})|=\frac{b_0}{2}$.
\item Otherwise, the less stable eigenvalues can be written as $s_\ell^+
= \frac{1}{2} (-b_0 + \sqrt{b_0^2-4k_0\lambda_\ell})$, which may be complex for some $\ell>1$. 
The least stable eigenvalue is obtained by setting
$\lambda_\ell=\lambda_1$, so that $S^{(\text{\tiny{RPAV}})}=|\Re(s_{\min})|=\frac{1}{2} (b_0 - \sqrt{b_0^2-4k_0\lambda_1})$. 
\end{enumerate}

For the case of RPRV feedback, following the same procedure as that of RPAV feedback, the characteristic equations  are given by
\begin{align}\label{eq:chara_eqn}
	s^2+\lambda_{\ell} b_0 s + \lambda_{\ell}  k_0=0.
\end{align}
For each $\ell \in \{1,2,\cdots, N\}$, the two roots of the characteristic equations~\eqref{eq:chara_eqn} are,
 \begin{align}\label{eq:eigenvalue_pos_vel1}
	 s_\ell^{\pm} =-\frac{\lambda_{\ell} b_0}{2}\pm \frac{\sqrt{(\lambda_{\ell}b_0)^2-4\lambda_{\ell}k_0}}{2}.
 \end{align}
Depending on the discriminant in~\eqref{eq:eigenvalue_pos_vel1}, there are three cases to analyze:
\begin{enumerate}
\item If $\lambda_N \leq 4k_0/b_0^2$, then the discriminant in~\eqref{eq:eigenvalue_pos_vel1} for each $\ell$ is non-positive. Recall that the stability margin is defined as the absolute value of the real part of the least stable eigenvalue, which yields 
\begin{align*}
	S^{(\text{\tiny{RPRV}})}=|\Re(s_{\min})|=\frac{\lambda_1 b_0}{2}.
\end{align*}
\item If $ \lambda_1 \geq 4k_0/b_0^2$, then the discriminant in~\eqref{eq:eigenvalue_pos_vel1} for each $\ell$ is non-negative, the \emph{less stable} eigenvalue can be written as
\begin{align*}
	s_\ell^+=-\frac{\lambda_{\ell} b_0-\sqrt{(\lambda_{\ell}b_0)^2-4\lambda_{\ell}k_0}}{2}= -\frac{2k_0}{b_0+\sqrt{b_0^2-4k_0/\lambda_\ell}}.
\end{align*}
The least stable eigenvalue is achieved by setting $\lambda_\ell=\lambda_N$, then have the stability margin
\begin{align*}
	S^{(\text{\tiny{RPRV}})}=|\Re(s_{\min})|=\frac{2k_0}{b_0+\sqrt{b_0^2-4k_0/\lambda_N}}.
\end{align*}
\item Otherwise, if the discriminant in~\eqref{eq:eigenvalue_pos_vel1}
  is negative for small $\ell$ and positive for large $\ell$, then the
  stability margin is given by taking the minimum of the two cases above. This completes the proof. \frQED
\end{enumerate}
}
\end{proof-lemma}

We are now ready to present the proof of Theorem~\ref{thm:sm_bound}.

\begin{proof-theorem}{\ref{thm:sm_bound}}
	We see from Lemma~\ref{lem:eig_relation} that the smallest and
        largest eigenvalues of matrix $L^{(1)}$ play important roles
        in determining the stability margin. To get a lower bound of
        the stability margin, a lower bound for the smallest
        eigenvalue and an upper bound for the largest eigenvalue is
        needed. Recall that $\lambda_1= 2 - 2\sqrt{1 - \epsilon^2}
        \cos \theta_1, \lambda_N=2 - 2\sqrt{1 - \epsilon^2} \cos
        \theta_N$, where $\theta_1 \in
        (\frac{\pi}{2(N+1)},\frac{3\pi}{2(N+1)}), \theta_N \in
        (\frac{(2N-1)\pi}{2(N+1)},\frac{(2N+1)\pi}{2(N+1)})$. We
        therefore have $\theta_1 \to 0, \theta_N \to \pi$  as $N \to
        \infty$, and consequently,
\begin{align}
	\inf_N \lambda_1&=2-2\sqrt{1-\epsilon^2}, \label{eq:inf} \\
	\sup_N \lambda_N&=2+2\sqrt{1-\epsilon^2}.\label{eq:sup} 
\end{align} 
\ifthenelse{\equal{\PaperORReport}{Paper}}{
To prove the result with RPAV feedback, we consider the
following two cases: (1) Case 1:  $\lambda_1 \geq b_0^2/4k_0$.  According to
Lemma~\ref{lem:eig_relation}, the stability margin is given by
$S^{(\text{\tiny{RPAV}})}=b_0/2$.  (2) Case 2: $\lambda_1 < b_0^2/4k_0$. From
Lemma~\ref{lem:eig_relation}, the stability margin is given by
\begin{align*}
	S^{(\text{\tiny{RPAV}})}=\frac{ b_0 - \sqrt{b_0^2-4k_0\lambda_1}}{2}.
	\end{align*}
Since $\lambda_1 \geq 2-2\sqrt{1-\epsilon^2}$, we obtain 
\begin{align}\label{eq:lb} 
	S^{(\text{\tiny{RPAV}})} \geq \frac{ b_0 - \sqrt{b_0^2-8k_0(1-\sqrt{1-\epsilon^2})}}{2}. 
\end{align}
Notice that the above lower bound~\eqref{eq:lb} is
smaller than $b_0/2$, the value of $S^{(\text{\tiny{RPAV}})}$ in case 1. The real part sign $\Re(.)$ in~\eqref{eq:sm_bound} comes from combining the above two cases. We obtain the first result of the theorem. 

The result for the RPRV feedback case again follows in a similar
manner, and an explicit proof is provided in~\cite{HH_PB_TAC_arxiv:11}.\frQED
}{
To prove the result with RPAV feedback, we consider the
following two cases: 
\begin{enumerate}
\item Case 1:  $\lambda_1 \geq b_0^2/4k_0$.  According to
Lemma~\ref{lem:eig_relation}, the stability margin is given by
$S^{(\text{\tiny{RPAV}})}=b_0/2$. 
\item Case 2: $\lambda_1 < b_0^2/4k_0$. From
Lemma~\ref{lem:eig_relation}, the stability margin is given by
\begin{align*}
	S^{(\text{\tiny{RPAV}})}=\frac{ b_0 - \sqrt{b_0^2-4k_0\lambda_1}}{2}.
	\end{align*}
Since $\lambda_1 \geq 2-2\sqrt{1-\epsilon^2}$, we obtain 
\begin{align}\label{eq:lb} 
	S^{(\text{\tiny{RPAV}})} \geq \frac{ b_0 - \sqrt{b_0^2-8k_0(1-\sqrt{1-\epsilon^2})}}{2}. 
\end{align}
\end{enumerate}
Notice that the above lower bound~\eqref{eq:lb} is
smaller than $b_0/2$, the value of $S^{(\text{\tiny{RPAV}})}$ in case 1. The real part sign $\Re(.)$ in~\eqref{eq:sm_bound} comes from combining the above two cases.  We obtain the first result of the theorem.

To prove the result with RPRV feedback, we consider the following three cases:
\begin{enumerate}
	\item Case 1: $\lambda_N \leq 4k_0/b_0^2$.  According to Lemma~\ref{lem:eig_relation}, the stability margin is  $S^{(\text{\tiny{RPRV}})}=b_0\lambda_1/2$. Moreover, from~\eqref{eq:inf}, we have $\inf_N \lambda_1=2-2\sqrt{1-\epsilon^2}$, therefore the stability margin has the lower bound 
\begin{align*}
	S^{(\text{\tiny{RPRV}})} \geq b_0(1-\sqrt{1-\epsilon^2}).
\end{align*}
	\item Case 2: $\lambda_1 \geq 4k_0/b_0^2$. From Lemma~\ref{lem:eig_relation}, the stability margin is given by 
\begin{align*}
	S^{(\text{\tiny{RPRV}})}=\frac{2k_0}{b_0+\sqrt{b_0^2-4k_0/\lambda_N}}.
	\end{align*}
In addition, we have  from~\eqref{eq:sup} that $\sup_N \lambda_N=2+2\sqrt{1-\epsilon^2}$, so the stability margin for this case is bounded below as 
\begin{align*} S^{(\text{\tiny{RPRV}})} \geq \frac{2k_0}{b_0+\sqrt{b_0^2-2k_0/(1+\sqrt{1-\epsilon^2})}}. \end{align*}
	\item Case 3: Otherwise, the stability margin are bounded below by the minimum of the above two cases. 
\end{enumerate}
Notice that in the second case, $\frac{2k_0}{b_0+\sqrt{b_0^2-2k_0/(1+\sqrt{1-\epsilon^2})}} \geq \frac{k_0}{b_0}$. Combining the above three cases, we have that 
\begin{align*}
	S^{(\text{\tiny{RPRV}})} \geq \min \Big \{ b_0(1-\sqrt{1-\epsilon^2}), \frac{k_0}{b_0}\Big \},
\end{align*}
which completes the proof. \frQED 
		}
\end{proof-theorem}

\subsection{Numerical verification for 1-D vehicular platoon}
In this section, we  present numerical verification of the lower bounds of the  stability
margins for both RPAV and RPRV feedbacks with asymmetric control, which are predicted by
Theorem~\ref{thm:sm_bound}. In addition, the stability margins with
symmetric control  are also computed to compare with the asymmetric
case. The stability margins are obtained by numerically evaluating the
eigenvalues of the state matrix $A^{(\text{\tiny{RPAV or RPRV}})}$
of~\eqref{eq:ss} with corresponding controllers. Figure~\ref{fig:fig3}
depicts the comparisons between the stability margins with symmetric
and asymmetric control for the two types of feedback: RPAV and
RPRV. For both symmetric and asymmetric controls, the nominal control
gains used are $k_0=1$, $b_0=0.5$, and for asymmetric control, the
amount of asymmetry is $\epsilon=0.1$.   We can see from
Figure~\ref{fig:fig3} that the stability margin of the vehicular
platoon with asymmetric control is indeed bounded away from $0$
uniformly in $N$, and the predictions Eq.~\eqref{eq:sm_bound} and
Eq.~\eqref{eq:sm_bound1} of Theorem~\ref{thm:sm_bound} are quite
accurate. Furthermore, for the same $N$, the stability margin with
asymmetric control is much larger than that with symmetric control, especially when $N$ is large. 
\begin{figure}
	    	\psfrag{N}{\small$N$}
		\psfrag{stability margin}{\small$S^{(\text{\tiny{RPAV or RPRV}})}$}
		\psfrag{S1}{\scriptsize$S^{(\text{\tiny{RPRV}})}$}
		\psfrag{sym}{\scriptsize Symmetric control ($\epsilon=0$)}
		\psfrag{asy}{\scriptsize Asymmetric control ($\epsilon=0.1$)}
		\psfrag{LB}{\scriptsize Lower bound in Theorem~\ref{thm:sm_bound}}
		\psfrag{s1}{\scriptsize Asymmetric RPAV}
		\psfrag{s21}{\scriptsize Symmetric RPAV}
		\psfrag{s3}{\scriptsize Asymmetric RPRV}
		\psfrag{s51}{\scriptsize Symmetric RPRV}
		\psfrag{lb1}{\scriptsize Eq.~\eqref{eq:sm_bound}}
		\psfrag{lb2}{\scriptsize Eq.~\eqref{eq:sm_bound1}}
		\psfrag{s2}{\scriptsize }
		\psfrag{s5}{\scriptsize }
\centering
\ifthenelse{\equal{\PaperORReport}{Paper}}{
\includegraphics[scale = 0.45]{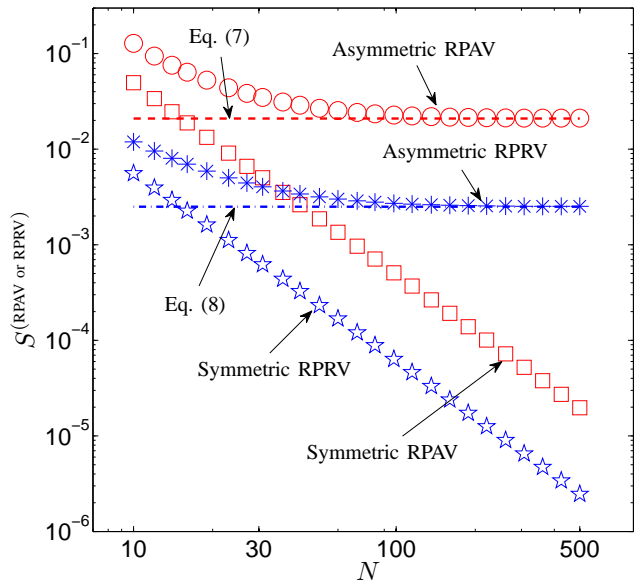} 
\caption{Stability margin comparisons between  symmetric control and asymmetric control.}
\label{fig:fig3}
\vspace{-0.8 cm}
\end{figure}
 }
 {
\includegraphics[scale = 0.45]{sm_combo.eps}
  \caption{Stability margin comparisons between  symmetric control and asymmetric control.}
\label{fig:fig3}
\end{figure}
}

\section{Stability margin with $D$-dimensional lattice information graph}\label{sec:sm_nD}
In this section we analyze a more general scenario than the 1-D
platoon of the previous sections. We consider 
a vehicular formation in which the position of each vehicle has dimension
higher than one, such as a vehicular formation moving in 2-D or 3-D space. We assume 
the dynamics of each of the coordinates of a vehicle's
position are decoupled and each coordinate can be
independently controlled. Under this \emph{fully actuated} assumption,
the closed loop dynamics for each coordinate of the position can be
independently studied; see~\cite{SD_PP_IJES:10,HH_PB_PM_TAC:11} for examples. The information used by a vehicle to compute its control is based on
relative measurements with a set of neighbors specified in
terms of an information graph. The problem formulation is similar to
the 1-D case in the sense that each vehicle has to maintain constant
separation with its neighbors in an information graph, except that the
information graph now is a $D$-dimensional lattice.  

\begin{definition} An \emph{information graph} is a graph $\G = (\V,\E)$, where the set of \emph{nodes} (vehicles) $\V=\{1,2,\dots, N, N+1, \dots, N + N_r\}$ consists of $N$ real vehicles and $N_r$ ``fictitious'' reference vehicles.  Two nodes $i$ and $j$ are called \emph{neighbors} if $(i,j) \in \E$, and the set of neighbors of $i$ are denoted by $\scr{N}_i$. \frqed
\end{definition}
 In this paper we restrict ourselves to $D$-dimensional  lattices as information graphs:
\begin{definition}[$D$-dimensional lattice] A $D$-dimensional  lattice,
  specifically a $n_1 \times n_2 \times \dots \times n_D$ lattice, is
  a graph with $n_1 n_2 \dots  n_D$ nodes, in which the nodes are placed at the integer coordinate points of the $D$-dimensional Euclidean space and each \emph{real} vehicle connects to vehicles which are exactly one unit away from it. \frqed
\end{definition}

\begin{figure}[h]
\begin{center}
\psfrag{x}{\scriptsize$\ X$}
\psfrag{x1}{\scriptsize$x_1$}
\psfrag{x2}{\scriptsize$x_2$}
\psfrag{o}{$o$}
\psfrag{4}{\scriptsize$6$} \psfrag{1}{\scriptsize$5$} \psfrag{5}{\scriptsize$4$} 
\psfrag{2}{\scriptsize$3$} \psfrag{6}{\scriptsize$2$} \psfrag{3}{\scriptsize$1$} 
\psfrag{7}{\scriptsize$7$} \psfrag{8}{\scriptsize$8$} \psfrag{9}{\scriptsize$9$} 
\psfrag{11}{\scriptsize$1$} \psfrag{12}{\scriptsize$2$} \psfrag{13}{\scriptsize$3$} 
\psfrag{14}{\scriptsize$4$} \psfrag{15}{\scriptsize$5$} \psfrag{16}{\scriptsize$6$} 
\psfrag{17}{\scriptsize$7$} \psfrag{18}{\scriptsize$8$} \psfrag{19}{\scriptsize$9$} 
\ifthenelse{\equal{\PaperORReport}{Paper}}{
\includegraphics[scale = 0.25]{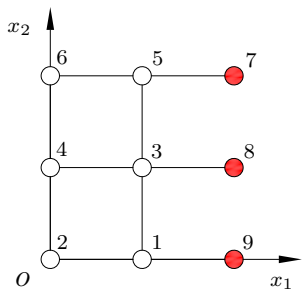} 
 }
 {
\includegraphics[scale = 0.25]{F1c.eps} 
 }
\end{center}
\caption{A pictorial representation of   a 2-D information graph. The filled node represent the reference vehicles and the solid lines represent edges in the information graph. }\label{fig:2D}
\end{figure}

Figure~\ref{fig:2D}  depicts an example of 2-D lattice.  A $D$-dimensional lattice is drawn in $\R^D$ with a Cartesian reference frame whose axes are denoted by $x_1,x_2,\dots,x_D$.  We also define $N_d \ (d=1,\dots,D)$  as the number of real vehicles in the $x_d$ direction. Then we have $N_1 N_2 \cdots N_D = N$ and $n_1 n_2 \dots n_D = N+N_r$. An information graph is said to be \emph{square} if $N_1=N_2=\cdots=N_D$. Note that the information graph for the vehicular platoon considered in the previous sections is a 1-D lattice with $N$ real vehicles (nodes) and $N_r\ (=1)$ reference vehicle. 

For the ease of exposition, we only consider the case where the
reference vehicles are arranged on one boundary of the
lattice. Without loss of generality, let it be perpendicular to the
$x_1$ axis, see Figure~\ref{fig:2D}  for an example.  This arrangement of
reference vehicles simplifies the presentation of the results. Arrangements of reference vehicles on other boundaries of the
lattice can also be considered, which does not significantly change
the results; see~\cite{HH_PB_JV_CDC:10,HH_PB_PM_DSCC:09}.

Due to its similarity with the 1-D case, we omit the details on
desired separations etc, which are available
in~\cite{HH_PB_PM_TAC:11}.  The control laws with RPAV and RPRV
feedback, in terms of the errors $\tilde{p}_i$ are, respectively
\ifthenelse{\equal{\PaperORReport}{Paper}}{
\begin{align}
	u_i=&-\sum_{d=1}^{D}   k_{(i,i^{d+})} (\tilde{p}_i - \tilde{p}_{i^{d+}})  -\sum_{d=1}^{D} k_{(i,i^{d-})}(\tilde{p}_i -   \tilde{p}_{i^{d-}}) \notag\\ &- b_i \dot{\tilde{p}}_i,\label{eq:control_law_1}\\
	u_i=&-\sum_{d=1}^{D}   k_{(i,i^{d+})} (\tilde{p}_i - \tilde{p}_{i^{d+}})  -\sum_{d=1}^{D} k_{(i,i^{d-})}(\tilde{p}_i -   \tilde{p}_{i^{d-}}) \notag\\ &-\sum_{d=1}^{D}   b_{(i,i^{d+})} (\dot{\tilde{p}}_i - \dot{\tilde{p}}_{i^{d+}})  -\sum_{d=1}^{D} b_{(i,i^{d-})}(\dot{\tilde{p}}_i -   \dot{\tilde{p}}_{i^{d-}}),\label{eq:control_law_2}
\end{align}}
{
\begin{align}
	u_i=&-\sum_{d=1}^{D}   k_{(i,i^{d+})} (\tilde{p}_i - \tilde{p}_{i^{d+}})  -\sum_{d=1}^{D} k_{(i,i^{d-})}(\tilde{p}_i -   \tilde{p}_{i^{d-}}) - b_i \dot{\tilde{p}}_i, \label{eq:control_law_1} \\
	u_i=&-\sum_{d=1}^{D}   k_{(i,i^{d+})} (\tilde{p}_i - \tilde{p}_{i^{d+}})  -\sum_{d=1}^{D} k_{(i,i^{d-})}(\tilde{p}_i -   \tilde{p}_{i^{d-}}) \notag\\ &-\sum_{d=1}^{D}   b_{(i,i^{d+})} (\dot{\tilde{p}}_i - \dot{\tilde{p}}_{i^{d+}})  -\sum_{d=1}^{D} b_{(i,i^{d-})}(\dot{\tilde{p}}_i -   \dot{\tilde{p}}_{i^{d-}}),\label{eq:control_law_2}
\end{align}
}
where $i^{d+}$ (respectively, $i^{d-}$) denotes the neighbor of $i$ on
the positive (respectively, negative) $x_d$ axis. The closed loop
dynamics are again represented as $\dot{x} = A^{(\text{\tiny{RPAV or RPRV}})}x$, where the state $x \eqdef
[\tilde{p}_1,\dot{\tilde{p}}_1,\cdots,\tilde{p}_N,\dot{\tilde{p}}_N] \in \R^{2N}$
is a vector of the relative positions $\tilde{p}_i$ and relative
velocities $\dot{\tilde{p}}_i$. The stability margin is defined as before. 

It is shown in~\cite{HH_PB_PM_TAC:11} that asymmetry in
control gains can improve the stability margin with RPAV feedback, but the analysis is limited for $\epsilon \to 0$ and the case with RPRV feedback was not considered. In this paper, we consider the
following homogeneous and asymmetric control gains that introduce asymmetry only in the $x_1$ axis: 
\begin{align}\label{eq:optimal-mistuned-gains}
\text{RPAV:  }	&
\begin{split}
& k_{(i,i^{1+})}=(1+\epsilon)k_0, \; k_{(i,i^{1-})}=(1-\epsilon)k_0, \\ 
& k_{(i,i^{d+})} = k_0, \ (d > 1),  \; b_i= b_0.  
\end{split}
\end{align}
\begin{align}\label{eq:optimal-mistuned-gains1}
\text{RPRV:  } &	
\begin{split}
	k_{(i,i^{1+})}=(1+\epsilon)k_0, & \quad k_{(i,i^{1-})}=(1-\epsilon)k_0, \\ b_{(i,i^{1+})}=(1+\epsilon)b_0, & \quad b_{(i,i^{1-})}=(1-\epsilon)b_0, \\ k_{(i,i^{d+})} = k_0 , &\quad  b_{(i,i^{d+})} = b_0, \ (d > 1).
\end{split}
\end{align}
Again, we comment that heterogeneity in control gains has little effect on the scaling trend of stability margin, not only for vehicular formation with lattice graphs but also for general graphs with bounded degree and weights, please refer to Lemma~\ref{lem:hete} given in the appendix.

We first summarize the results in~\cite{HH_PB_PM_TAC:11, HH_PB_JV_CDC:10}.
\begin{proposition}\label{prop:summary1}
	Consider a vehicular formation  whose information graph is a $D$-dimensional lattice. With the control gains given in~\eqref{eq:optimal-mistuned-gains} and~\eqref{eq:optimal-mistuned-gains1} respectively.
        \begin{enumerate}
        \item [1)] [Theorem 1 of~\cite{HH_PB_PM_TAC:11}, Theorem 4 of~\cite{HH_PB_JV_CDC:10}] With symmetric control ($\epsilon=0$), both $S^{(\text{\tiny{RPAV}})}$ and $S^{(\text{\tiny{RPRV}})}$ are $O(\frac{1}{N_1^2}).$
      \item [2)] [Theorem 2 of~\cite{HH_PB_PM_TAC:11}] With the control gains given
        by~\eqref{eq:optimal-mistuned-gains},  the 
         stability margin with RPAV feedback is
         $S^{(\text{\tiny{RPAV}})}  = O(\frac{\epsilon}{N_1})$, which
         hold in the limit $\epsilon \to 0$ and $N_1 \to \infty$. \frqed
        \end{enumerate}
\end{proposition}

We next state the main result of this section, which is a corollary of Theorem~\ref{thm:sm_bound}. It describes the
stability margin for a vehicular formation with $D$-dimensional lattice information graph with asymmetric control. 
\begin{corollary}\label{cor:sm_bound_nD}
	With the control gains given in~\eqref{eq:optimal-mistuned-gains} and~\eqref{eq:optimal-mistuned-gains1} respectively,  and $0<\epsilon<1$, the stability margin of the vehicular formation with RPAV or RPRV feedback is bounded away from $0$, uniformly in $N$. Specifically,
\ifthenelse{\equal{\PaperORReport}{Paper}}{
\begin{align*}
S^{(\text{\tiny{RPAV}})} \geq & \frac{\Re \Big (b_0-\sqrt{b_0^2-8k_0(1-\sqrt{1-\epsilon^2})}\Big )}{2},\\
S^{(\text{\tiny{RPRV}})} \geq & \min \Big \{ b_0(1-\sqrt{1-\epsilon^2}), \ \frac{k_0}{b_0} \Big\}. \tag*{\frqed} 
\end{align*}

\smallskip

}
{
\begin{align}
	S^{(\text{\tiny{RPAV}})} \geq & \frac{\Re \Big (b_0-\sqrt{b_0^2-8k_0(1-\sqrt{1-\epsilon^2})}\Big )}{2},\label{eq:sm_bound_nD}\\
S^{(\text{\tiny{RPRV}})} \geq & \min \Big \{ b_0(1-\sqrt{1-\epsilon^2}), \ \frac{k_0}{b_0} \Big\}. \label{eq:sm_bound_nD1} 
\end{align} \frqed
}
\end{corollary} 

\begin{remark}
  From Proposition~\ref{prop:summary1}, we see that with the particular
  arrangement of the reference vehicles as mentioned before, the
  stability margin of the vehicular formation with symmetric control
   only depend on $N_1$, the number of real vehicles along the $x_1$
  axis of the information graph. For a \emph{square} information graph,
  no matter how large its dimension $D$ is, the loss of stability
  margin with increasing number of vehicle $N$ is inevitable, since
  $N_1 = N^{1/D}$. To make the stability margin independent of $N$
  with symmetric control,
  one needs to employ a non-square information graph, such that $N_1$
  is a constant regardless of the increasing of $N$. The price one
  pays is either long range communication and/or increased number of
  reference vehicles; see~\cite{HH_PB_PM_TAC:11,HH_PB_JV_CDC:10} for
  more details. In addition, for the RPAV feedback case, with vanishingly small amount of asymmetry, the stability margin is improved to $O(1/N_1)$,  compared to the $O(1/N_1^2)$ trend in the symmetric case.
 
 In contrast, Corollary~\ref{cor:sm_bound_nD} shows that with judicious asymmetric
  control, the stability
  margin can be made  independent of the number of vehicles $N$ in
  the formation, without using the non-square information graph aforementioned. Note that the result we establish in this paper
  (Corollary~\ref{cor:sm_bound_nD}) is stronger than that
  in~\cite{HH_PB_PM_TAC:11}, even though the  control law is the same. The reason is
  that the analysis in~\cite{HH_PB_PM_TAC:11} relied on a perturbation
  technique, which limited its applicability to   vanishingly small
  $\epsilon$. In this paper we do not use perturbation
  techniques, and obtain result for any non-vanishing $\epsilon \in (0,1)$. In
  addition, we also consider the RPRV feedback case,
  while~\cite{HH_PB_PM_TAC:11} analyzed only RPAV feedback. \frqed
\end{remark}

\begin{proof-corollary}{\ref{cor:sm_bound_nD}}
With the control gains specified in~\eqref{eq:optimal-mistuned-gains} and~\eqref{eq:optimal-mistuned-gains1} respectively, it is straightforward - through a bit tedious - to show that the state matrices $A^{(\text{\tiny{RPAV}})}$ and $A^{(\text{\tiny{RPRV}})}$ can be expressed in the following forms,
\begin{align}\label{eq:matrix_relation}
	A^{(\text{\tiny{RPAV}})} = I_N \otimes A_1 + L^{(D)} \otimes A_2, \notag\\
	A^{(\text{\tiny{RPRV}})} = I_N \otimes A_3 + L^{(D)} \otimes A_4,
\end{align}
where $A_1, A_2, A_3, A_4$ are given in~\eqref{eq:auxi_matrices} and $L^{(D)}$ has the following form:
\begin{align}\label{eq:DD_induction}
	L^{(d)} = I_{N_d} \otimes L^{(d-1)}+ T^{(d)} \otimes I_{N_1 N_2\cdots N_{d-1}}, \quad 2 \leq d \leq D,
\end{align}
where $L^{(1)}$ is given in~\eqref{eq:redu_laplacian} and $T^{(d)}$ is a matrix of dimension $N_d \times N_d$, which is given by
\begin{align}\label{eq:triangular3}
	T^{(d)}=\begin{bmatrix}
		1 & -1&  &   & \\
		-1 & 2& -1 &  &  \\
		  & \ddots &  \ddots & \ddots &\\
		  &    & -1 & 2 & -1\\
		  &     &  & -1 & 1
	       \end{bmatrix}.
\end{align}
The eigenvalues of $T^{(d)}$ are given by (see~\cite{yueh2008explicit}):
\begin{align} \label{eq:eig3}
\lambda_{\ell_d}=2-2\cos \frac{(\ell_d-1) \pi}{N_d}, \quad \ell_d=1,2,\dots,N_d.
\end{align}
\ifthenelse{\equal{\PaperORReport}{Paper}}{
}
{
For example, for a $2$-dimensional information graph shown in Figure~\ref{fig:2D}, 
\begin{align*}
	L^{(2)} & =  \begin{bmatrix}
                3 & -1+\epsilon& -1 & 0 & 0 & 0\\
		-1-\epsilon & 2+\epsilon& 0 & -1 & 0 & 0\\
		-1 & 0 & 4& -1 +\epsilon& -1 & 0 \\
		0 & -1 & -1-\epsilon & 3+\epsilon& 0 & -1  \\
		0 & 0& -1 & 0 & 3 & -1+\epsilon\\
		0 & 0& 0 & -1 & -1 -\epsilon& 2+\epsilon \end{bmatrix}. 
\end{align*}
It's straightforward to show that $L^{(2)} = I_{3} \otimes L^{(1)}+ T^{(2)} \otimes I_{2}$,
where $T^{(2)}$ is a matrix with dimension $3 \times
3$. }

From the proof of Lemma~\ref{lem:eig_relation}, we see that the
eigenvalues of $A^{(\text{\tiny{RPAV}})}$ and $A^{(\text{\tiny{RPRV}})}$ are given by the roots  of the 
characteristic equations $s^2+b_0s+k_0\lambda_{\vec{\ell}}=0$ and $s^2+b_0\lambda_{\vec{\ell}}s+k_0\lambda_{\vec{\ell}}=0$ respectively, where $\lambda_{\vec{\ell}}$ is the eigenvalue of $L^{(D)}$, and  $\vec{\ell}=(\ell_1,\cdots,\ell_D)$ in which $\ell_d \in \{1,2,\cdots, N_d\}$.
We next claim that the eigenvalues of $L^{(D)}$ are given by 
\begin{align}
	\lambda_{\vec{\ell}}=\lambda_{\ell_1}(L^{(1)}) +\sum_{d=2}^{D} \lambda_{\ell_d} (T^{(d)}).
\end{align}
We prove by induction method. For the case $d=2$, $L^{(2)}=I_{N_2}\otimes L^{(1)}+T^{(2)}\otimes I_{N_1}$.  
Following~\eqref{eq:eigen_relation} in the proof of Lemma~\ref{lem:eig_relation}, the eigenvalues of $L^{(2)}$ are given by
\begin{align*}
	\lambda_{\ell_1,\ell_2} &=  \bigcup_{\lambda_{\ell_2} \in \sigma(T^{(2)})} \{ \sigma (L^{(1)}+\lambda_{\ell_2} I_{N_1})\}  \\&=\lambda_{\ell_1}(L^{(1)})+\lambda_{\ell_2}(T^{(2)}), 
\end{align*}
Now, we assume the general formula for the eigenvalues of  $L^{(D-1)}$ is given by 
\begin{align}\label{eq:eig_D1D_all}
\lambda_{\ell_1, \dots, \ell_{D-1}} = \lambda_{\ell_1}(L^{(1)}) +\sum_{d=2}^{D-1} \lambda_{\ell_d} (T^{(d)}) .
\end{align}
For the case $d=D$, the matrix $L^{(D)}$ has the form given in~\eqref{eq:DD_induction}, use~\eqref{eq:eigen_relation} again, we have
\begin{align*}
	\lambda_{\ell_1, \dots, \ell_{D}} &=  \bigcup_{\lambda_{\ell_D} \in \sigma(T^{(D)})} \{ \sigma (L^{(D-1)}+\lambda_{\ell_D} I_{N_1\cdots N_{D-1}})\} \\&=\lambda_{\ell_1 \cdots \ell_{D-1}}(L^{(D-1)})+\lambda_{\ell_D}(T^{(D)}),
\end{align*}
which proves the claim. Now, use~\eqref{eq:lambda_formula}
and~\eqref{eq:eig3}, the smallest eigenvalue of $L^{(D)}$ is equal to
$\lambda_1$, the smallest eigenvalue of $L^{(1)}$. The result now
follows from Lemma~\ref{lem:eig_relation} and Theorem~\ref{thm:sm_bound}.\frQED
\end{proof-corollary}

\ifthenelse{\equal{\PaperORReport}{Paper}}{ Numerical verification is
  omitted here due to lack of space; it is available in~\cite{HH_PB_TAC_arxiv:11}. } {
\subsection{Numerical verification for vehicular formation with D-dimensional information graph}
In this section, we present numerical verification of the theoretical predicted lower bounds of  stability
margin for vehicular formations with D-dimensional lattice information
graphs. For simplicity, we take $2$-D lattices as examples. We assume the information graph is \emph{square},
i.e. $N_1=N_2=\sqrt{N}$. In addition, the stability margins with
symmetric control  are also computed to compare with the asymmetric
case. For both symmetric and asymmetric controls, the nominal control
gains used are $k_0=1$, $b_0=0.5$, and for asymmetric control, the
amount of asymmetry used is $\epsilon=0.1$.  We observe from
Figure~\ref{fig:fig4} that, with asymmetric control, the stability margin of the vehicular
formation with RPAV or RPRV feedback  is indeed uniformly bounded below by
the prediction Eq.~\eqref{eq:sm_bound_nD} and
Eq.~\eqref{eq:sm_bound_nD1}  respectively. Furthermore the stability margin with
asymmetric control is much larger than that with symmetric control for the same $N$. 

\begin{figure}
	    	\psfrag{N}{\small$N$}
		\psfrag{stability margin}{\small$S^{(\text{\tiny{RPAV or RPRV}})}$}
		\psfrag{S1}{\scriptsize$S^{(\text{\tiny{RPRV}})}$}
		\psfrag{sym}{\scriptsize Symmetric control ($\epsilon=0$)}
		\psfrag{asy}{\scriptsize Asymmetric control ($\epsilon=0.1$)}
		\psfrag{LB}{\scriptsize Lower bound in Theorem~\ref{thm:sm_bound}}
		\psfrag{s1}{\scriptsize Asymmetric RPAV}
		\psfrag{s21}{\scriptsize Symmetric RPAV}
		\psfrag{s3}{\scriptsize Asymmetric RPRV}
		\psfrag{s51}{\scriptsize Symmetric RPRV}
		\psfrag{lb1}{\scriptsize Eq.~\eqref{eq:sm_bound_nD}}
		\psfrag{lb2}{\scriptsize Eq.~\eqref{eq:sm_bound_nD1}}
		\psfrag{s2}{\scriptsize }
		\psfrag{s5}{\scriptsize }
\centering
\includegraphics[scale = 0.45]{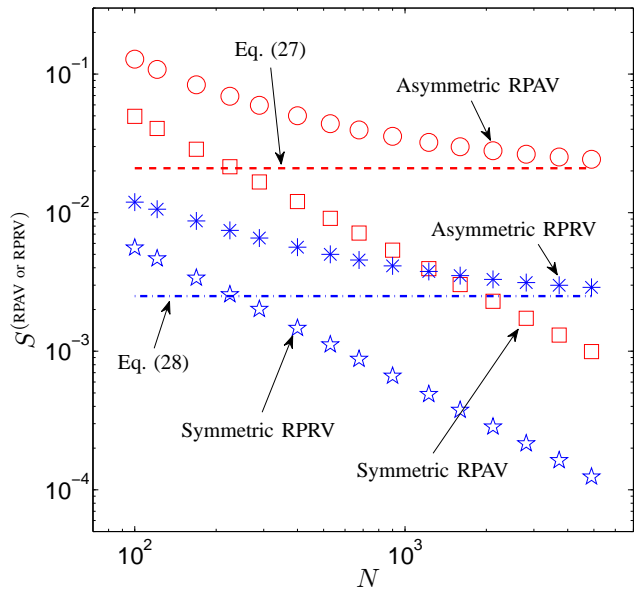}
 
  \caption{Stability margin comparisons between symmetric control and asymmetric control for a vehicular formation with 2-D square information graph.}
\label{fig:fig4}
\end{figure}
}

\section{Summary}\label{sec:conc}

We studied the stability margin of vehicular formations on lattice
graphs with distributed control.  The control signal at every vehicle
depends on the measurements from its neighbors in the information
graph, which is a $D$-dimensional lattice. Inspired by the previous
works~\cite{PB_PM_JH_TAC:09,HH_PB_PM_TAC:11}, we examined the role of
asymmetry in the control gains on the closed loop stability margin. We
showed that with judicious asymmetry in the control gains, the
stability margin of the vehicular formation can be bounded away from
$0$ uniformly in $N$. This eliminates the loss of stability margin
with increasing $N$ that is seen with symmetric control. In this
paper, the analysis of the stability margin avoids the PDE
approximation and perturbation method used
in~\cite{PB_PM_JH_TAC:09,HH_PB_PM_TAC:11}. In particular, the latter
limited the analyses in those papers to vanishingly small amount of
asymmetry and resulted a $O(1/N)$ scaling trend of stability
margin. In addition, the control laws examined
in~\cite{PB_PM_JH_TAC:09,HH_PB_PM_TAC:11} required vehicles to have
access to the desired velocity of the formation. We generalized the
results to the case when only relative velocity and relative position measurements are
available. We showed in this paper that in both cases (i.e., with or without
absolute velocity feedback), stability margin can be made independent
of the size of the formation with asymmetric
control. The issue of sensitivity to external disturbances with
asymmetric control  is a topic of future research. 

\bibliographystyle{IEEEtran}        
\bibliography{../../../PBbib/vehicular_platoon,../../../PBbib/sensnet_bib_dbase,../../../PBbib/Barooah,../../../PBbib/HH,../../../PBbib/distributed_control}

\begin{thebibliography}{10}
\providecommand{\url}[1]{#1}
\csname url@rmstyle\endcsname
\providecommand{\newblock}{\relax}
\providecommand{\bibinfo}[2]{#2}
\providecommand\BIBentrySTDinterwordspacing{\spaceskip=0pt\relax}
\providecommand\BIBentryALTinterwordstretchfactor{4}
\providecommand\BIBentryALTinterwordspacing{\spaceskip=\fontdimen2\font plus
\BIBentryALTinterwordstretchfactor\fontdimen3\font minus
  \fontdimen4\font\relax}
\providecommand\BIBforeignlanguage[2]{{%
\expandafter\ifx\csname l@#1\endcsname\relax
\typeout{** WARNING: IEEEtran.bst: No hyphenation pattern has been}%
\typeout{** loaded for the language `#1'. Using the pattern for}%
\typeout{** the default language instead.}%
\else
\language=\csname l@#1\endcsname
\fi
#2}}

\bibitem{PB_PM_JH_TAC:09}
P.~Barooah, P.~G. Mehta, and J.~P. Hespanha, ``Mistuning-based decentralized
  control of vehicular platoons for improved closed loop stability,''
  \emph{{IEEE} Transactions on Automatic Control}, vol.~54, no.~9, pp.
  2100--2113, September 2009.

\bibitem{HH_PB_PM_TAC:11}
H.~Hao, P.~Barooah, and P.~G. Mehta, ``Stability margin scaling laws of
  distributed formation control as a function of network structure,''
  \emph{{IEEE} Transactions on Automatic Control}, vol.~56, pp. 923 -- 929,
  April 2011.

\bibitem{JH_MT_PV_CSM:94}
J.~K. Hedrick, M.~Tomizuka, and P.~Varaiya, ``Control issues in automated
  highway systems,'' \emph{{IEEE} Control Systems Magazine}, vol.~14, pp. 21 --
  32, December 1994.

\bibitem{okubo_flock}
A.~Okubo, ``{Dynamical aspects of animal grouping: swarms, schools, flocks, and
  herds},'' \emph{Advances in Biophysics}, vol.~22, pp. 1--94, 1986.

\bibitem{DJ_WB_MP_EW_AIAA:02}
E.~Wagner, D.~Jacques, W.~Blake, and M.~Pachter, ``Flight test results of close
  formation flight for fuel savings,'' in \emph{{AIAA} Atmospheric Flight
  Mechanics Conference and Exhibit}, 2002, {AIAA}-2002-4490.

\bibitem{SD_PP_IJES:10}
S.~Darbha and P.~R. Pagilla, ``Limitations of employing undirected information
  flow graphs for the maintenance of rigid formations for heterogeneous
  vehicles,'' \emph{International journal of engineering science}, vol.~48,
  no.~11, pp. 1164--1178, 2010.

\bibitem{das2002framework}
A.~Das, R.~Fierro, V.~Kumar, J.~Ostrowski, J.~Spletzer, and C.~Taylor, ``A
  framework for vision based formation control,'' \emph{IEEE Transactions on
  Robotics and Automation}, vol.~18, no.~5, pp. 813--825, 2002.

\bibitem{tanner2004leader}
H.~Tanner, G.~Pappas, and V.~Kumar, ``Leader-to-formation stability,''
  \emph{Robotics and Automation, IEEE Transactions on}, vol.~20, no.~3, pp.
  443--455, 2004.

\bibitem{stringstability:96}
S.~Darbha and J.~K. Hedrick, ``String stability of interconnected systems,''
  \emph{IEEE Transactions on Automatic Control}, vol.~41, no.~3, pp. 349--356,
  March 1996.

\bibitem{Seiler_disturb_TAC:04}
P.~Seiler, A.~Pant, and J.~K. Hedrick, ``Disturbance propagation in vehicle
  strings,'' \emph{IEEE Transactions on Automatic Control}, vol.~49, pp.
  1835--1841, October 2004.

\bibitem{bamjovmitCDC08}
B.~Bamieh, M.~R. Jovanovi\'c, P.~Mitra, and S.~Patterson, ``Effect of
  topological dimension on rigidity of vehicle formations: fundamental
  limitations of local feedback,'' in \emph{Proceedings of the 47th IEEE
  Conference on Decision and Control}, Cancun, Mexico, 2008, pp. 369--374.

\bibitem{illposed}
M.~R. Jovanovi\'c and B.~Bamieh, ``On the ill-posedness of certain vehicular
  platoon control problems,'' \emph{IEEE Trans. Automatic Control}, vol.~50,
  no.~9, pp. 1307--1321, September 2005.

\bibitem{veerman2010asymmetric}
\BIBentryALTinterwordspacing
F.~Tangerman and J.~Veerman, ``{Asymmetric Decentralized Flocks},''
  \emph{accepted to IEEE Transactions on Automatic Control}, 2011. [Online].
  Available: \url{http://www.mth.pdx.edu/~veerman/publ04.html}
\BIBentrySTDinterwordspacing

\bibitem{HH_PB_CDC:10}
H.~Hao and P.~Barooah, ``Control of large {1D} networks of double integrator
  agents: role of heterogeneity and asymmetry on stability margin,'' in
  \emph{{IEEE} Conference on Decision and Control}, December 2010.

\bibitem{veerman_automated}
J.~Veerman, B.~Sto{\v{s}}i{\'c}, and F.~Tangerman, ``{Automated traffic and the
  finite size resonance},'' \emph{Journal of Statistical Physics}, vol. 137,
  no.~1, pp. 189--203, October 2009.

\bibitem{SwaroopHedrick_stringstability_TAC:96}
S.~Darbha and J.~K. Hedrick, ``String stability of interconnected systems,''
  \emph{IEEE Transactions on Automatic Control}, vol.~41, no.~3, pp. 349--356,
  March 1996.

\bibitem{zhang1999using}
Y.~Zhang, B.~Kosmatopoulos, P.~Ioannou, and C.~Chien, ``{Using front and back
  information for tight vehicle following maneuvers},'' \emph{IEEE Transactions
  on Vehicular Technology}, vol.~48, no.~1, pp. 319--328, 1999.

\bibitem{Middleton_TAC:09}
R.~Middleton and J.~Braslavsky, ``{String instability in classes of linear time
  invariant formation control with limited communication range},'' \emph{IEEE
  Transactions on Automatic Control}, vol.~55, no.~7, pp. 1519--1530, 2010.

\bibitem{darbha1994comparison}
S.~Darbha, J.~Hedrick, C.~Chien, and P.~Ioannou, ``A comparison of spacing and
  headway control laws for automatically controlled vehicles,'' \emph{Vehicle
  System Dynamics}, vol.~23, no.~8, pp. 597--625, 1994.

\bibitem{feedback_linearization}
S.~Stankovic, M.~Stanojevic, and D.~Siljak, ``Decentralized overlapping control
  of a platoon of vehicles,'' \emph{Control Systems Technology, IEEE
  Transactions on}, vol.~8, no.~5, pp. 816--832, 2000.

\bibitem{yueh2008explicit}
W.~Yueh and S.~Cheng, ``Explicit eigenvalues and inverses of tridiagonal
  toeplitz matrices with four perturbed corners,'' \emph{The Australian \& New
  Zealand Industrial and Applied Mathematics (Anziam) Journal}, vol.~49, no.~3,
  pp. 361--388, 2008.

\bibitem{veerman2005flocks}
J.~Veerman, G.~Lafferriere, J.~Caughman, and A.~Williams, ``{Flocks and
  formations},'' \emph{Journal of Statistical Physics}, vol. 121, no.~5, pp.
  901--936, 2005.

\bibitem{HH_PB_JV_CDC:10}
H.~Hao, P.~Barooah, and J.~J.~P. Veerman, ``Effect of network structure on the
  stability margin of vehicle formation with distributed control,'' in
  \emph{{IEEE} Conference on Decision and Control}, December 2010, pp. 4783 --
  4788.

\bibitem{HH_PB_PM_DSCC:09}
H.~Hao, P.~Barooah, and P.~G. Mehta, ``Distributed control of two dimensional
  vehicular formations: stability margin improvement by mistuning,'' in
  \emph{{ASME} Dynamic Systems and Control Conference}, October 2009, pp.
  699--706.

\bibitem{SKY_SD_KRR_TAC:06}
S.~K. Yadlapalli, S.~Darbha, and K.~R. Rajagopal, ``Information flow and its
  relation to stability of the motion of vehicles in a rigid formation,''
  \emph{{IEEE} Transactions on Automatic Control}, vol.~51, no.~8, August 2006.

\bibitem{barooah2006graph}
P.~Barooah and J.~Hespanha, ``Graph effective resistance and distributed
  control: Spectral properties and applications,'' in \emph{Decision and
  Control, 2006 45th IEEE Conference on}.\hskip 1em plus 0.5em minus
  0.4em\relax IEEE, 2006, pp. 3479--3485.

\bibitem{khalil}
H.~Khalil, \emph{{Nonlinear Systems 3rd}}.\hskip 1em plus 0.5em minus
  0.4em\relax Prentice hall Englewood Cliffs, NJ, 2002.

\end{thebibliography}

\appendix

\begin{lemma}\label{lem:hete}
Consider a vehicular formation whose information graph is an arbitrary connected graph (with $1$ grounded node/leader) with bounded degree and weights. With the double integrator dynamics~\eqref{eq:vehicle-dynamics} and the following heterogeneous and symmetric control gains with RPAV and RPRV feedback  respectively, 
\begin{align*}
	u_i^{(\text{\tiny{RPAV}})}=&-\sum_{j \in \scr{N}_i}   k_0w_{i,j} (\tilde{p}_i - \tilde{p}_{j})  - b_0 \dot{\tilde{p}}_i,\label{eq:control_law_1}\\
	u_i^{(\text{\tiny{RPRV}})}=&-\sum_{j \in \scr{N}_i}   k_0w_{i,j} (\tilde{p}_i - \tilde{p}_{j})  -\sum_{j \in \scr{N}_i}   b_0 w_{i,j} (\dot{\tilde{p}}_i - \dot{\tilde{p}}_{j}),
\end{align*}
where $k_0,b_0$ are positive constants and $w_{i,j}=w_{j,i}>0$ are the weight on the edge $(i,j)$ and $(j,i)$ respectively. The stability margin decays to zero as $N$ goes to infinity, i.e.
	\begin{align*}
S^{(\text{\tiny{RPAV or RPRV}})}  \to 0 \qquad \text{as} \qquad N \to \infty,
\end{align*}
where $N$ is the number of vehicles in the formation. \frqed

\end{lemma}
\begin{proof-lemma}{\ref{lem:hete}}
The state matrices $A^{(\text{\tiny{RPAV}})}$ and $A^{(\text{\tiny{RPRV}})}$ can be expressed in the following forms,
\begin{align*}
	A^{(\text{\tiny{RPAV}})} = I_N \otimes A_1 + L \otimes A_2, \notag\\
	A^{(\text{\tiny{RPRV}})} = I_N \otimes A_3 + L \otimes A_4,
\end{align*}
where $A_1, A_2, A_3, A_4$ are given in~\eqref{eq:auxi_matrices} and $L$ is the grounded graph Laplacian of the vehicular formation. First, we recall that the Laplacian matrix $\scr{L}$ of an arbitrary graph $\G$ with edge weights $w_{i,j}$ is defined as
\begin{align*}
	\scr{L}_{i,j}=
	\begin{cases}  	-w_{i,j} & i \neq j, (i,j) \in \E ,\\
		\sum_{k=1}^{N} w_{i,k} & i=j, (i,k) \in \E,\\
			0&\text{ otherwise.}
			\end{cases}
\end{align*}
The grounded graph Laplacian $L$ is obtained by deleting the row and column of $\scr{L}$ corresponding to the leader (recall that we have a leader/reference vehicle). Without loss of generality, let it be indexed by $1$. The neighbors of the leader is denoted by $\scr{N}_1$.

From the proof of Lemma~\ref{lem:eig_relation}, we notice that the stability margin of the formation is determined by $\lambda_1$, the smallest eigenvalue of its grounded graph Laplacian $L$. 

We first claim that as long as its grounded graph Laplacian $L$ is symmetric, i.e.  $w_{i,j}=w_{j,i}$, the smallest eigenvalue $\lambda_1$ of $L$ satisfies 
	\begin{align*}
		\lambda_1 \to 0 \qquad \text{as} \qquad N \to \infty.
	\end{align*}
This fact can be seen as a generalization of Lemma $3.2$ of~\cite{SKY_SD_KRR_TAC:06}.	First of all, the grounded graph Laplacian $L$ is positive definite~\cite{barooah2006graph}. From the Rayleigh Ritz Theorem~\cite{khalil}, for an arbitrary vector $x$, we have the following inequality 
\begin{align*}
	x^T \lambda_1(L) x \leq x^T L x, \qquad \Rightarrow \qquad \lambda_1 \leq \frac{x^T L x}{x^T x}.
\end{align*}
In particular, we pick the following vector with dimension $(N-1) \times 1$, $x=[1,1, \cdots, 1]^T$, then we have 
\begin{align*}
	\lambda_1 \leq \frac{x^T L x}{N-1}\leq \frac{|\scr{N}_1|\max_{j\in \scr{N}_1} w_{j,1}}{N-1},
\end{align*}
where $|\scr{N}_1|$ denotes the number of neighbors of the leader. Since the weight $w_{j,1}$ and $|\scr{N}_1|$ are bounded, when $N$ goes to infinity, the fact follows. With this, the rest of the proof follows from Lemma~\ref{lem:eig_relation}.	 \frQED

\end{proof-lemma}
\end{document}